\documentclass[12pt]{article}
\usepackage[english]{babel}
\newcommand{\bd}{\begin{displaymath}}
\newcommand{\ed}{\end{displaymath}}
\newcommand{\be}{\begin{equation}}
\newcommand{\ee}{\end{equation}}
\newcommand{\beq}{\begin{eqnarray}}
\newcommand{\eeq}{\end{eqnarray}}  
\newcommand{\beqs}{\begin{eqnarray*}}
\newcommand{\eeqs}{\end{eqnarray*}} 
\begin{document}
\begin{center}
\huge{Perturbations of planar interfaces in Ginzburg-Landau models}
\end{center}
\begin{center}
\Large{by}
\end{center}
\begin{center}
\LARGE{ H. Arod\'z, R. Pe\l{}ka and \L{}. St\c{e}pie\'n }
\end{center}
\begin{center}
\Large{Smoluchowski Institute of Physics, \\
Jagellonian University, \\   Reymonta 4, 30-059 Cracow, Poland}
\end{center}

\vspace*{2cm}

Certain dissipative Ginzburg-Landau models predict existence of planar 
interfaces moving with constant velocity. In most cases the interface 
solutions are hard 
to obtain because pertinent evolution equations are nonlinear. We present
a systematic perturbative expansion which allows us to compute effects of
small terms added to the free energy functional of a soluble model. 
As an example,
we take the exactly soluble model with single order parameter $\varphi$ 
and the potential $V_0(\varphi) = A\varphi^2 + B \varphi^3 + \varphi^4$, 
and we perturb it by adding
$V_1(\phi) = \frac{1}{2}\varepsilon_1\varphi^2 \partial_i\varphi 
\partial_i\varphi
+\frac{1}{5}\varepsilon_2 \varphi^5 + \frac{1}{6}\varepsilon_3 \varphi^6. $
We discuss the corresponding changes of the velocity of the planar interface.

\vspace*{2cm}

TPJU-1/2001

\newpage
\section{Introduction}
\indent 
Properties and evolution of interfaces is a very interesting topic in 
soft condensed matter physics, see, e. g., \cite{1} for a recent 
review. There are 
two main lines of theoretical research in that direction. The first one,
computer simulations, provides a relation to microscopic, molecular dynamics
level \cite{2,3}. In the second approach one is satisfied with a 
coarse-grained description by effective Ginzburg-Landau models. 

In the framework of Ginzburg-Landau models the interfaces are described by 
particular solutions of pertinent 
nonlinear evolution equations. The nonlinearity is crucial for 
existence of these solutions, but it is also a serious obstacle in obtaining 
them in an explicit form. In most cases only numerical 
solutions are available. A well-known exception is a model which appears 
in the context of reaction-diffusion chemical processes \cite{4}, and also is 
related to Landau-deGennes effective model for nematic liquid crystals 
\cite{5}. It involves a single, non-conserved, real-valued order parameter
$\varphi(x^{\alpha},t)$, and the free energy functional has the form 
\be
F_0=\int d^3 x\left(\frac{1}{2}\nabla\varphi\nabla\varphi+V_0(\varphi)\right),
\ee
\be V_0(\varphi)=A{\varphi}^2+B{\varphi}^3+{\varphi}^4, 
\ee 
where 
$\nabla=\left(\frac{\partial}{\partial x^{\alpha}}\right)$, $\alpha=1,2,3$,
$x^{\alpha}$ are Cartesian coordinates in the space.
For simplicity we have rescaled $x^{\alpha}, \varphi$ and $F_0$ 
so that they are dimensionless and the coefficient 
in front of the quartic 
term in $V_0(\varphi)$ is equal to $1$. The constants $A,B$ are positive, and
$A$ is restricted to the interval $0<A<\frac{9B^2}{32}$. 
Then $V(\varphi)$ has two minima and a maximum between them. 
In this model exact formulas for profile and velocity of a planar interface 
are known, see, e.g., \cite{4}.  

Exactly soluble models provide natural starting points for perturbative
expansions. If the expression (1) for the free energy is
changed a little bit, one may hope that the corresponding change of the 
planar interface can be calculated with the help of a perturbative expansion 
around the known solution. In this way we would obtain analytical results 
for velocity and profile of planar interfaces in a class of Ginzburg-Landau 
models. Moreover, this would open the way for investigations of evolution
of curved interfaces in that class of models, because one could use the
expansion in curvatures \cite{6} for which the knowledge of planar interfaces
is the only prerequisite.

In the present paper we explore the idea of the perturbative expansion 
for the planar interfaces. We show that indeed such perturbative scheme can
be constructed. Corrections to the profile are obtained in it as
solutions of ordinary, linear differential equations, while corrections to 
the velocity follow from integrability conditions which are due to existence 
of zero-modes. This type of
integrability conditions is well-known in statistical physics \cite{7}. 
As an example we consider perturbations of the form
\bd \delta F=\int d^3 x V_1  ,\ed
where
\be
V_1=\frac{1}{2} 
\varepsilon_1{\varphi}^2\nabla\varphi\nabla\varphi+
\frac{1}{5}\varepsilon_2{\varphi}^5+\frac{1}{6}\varepsilon_3{\varphi}^6.
\ee
Values of the perturbation parameters $\varepsilon_1, \varepsilon_2, 
\varepsilon_3$ are such 
that $F_0+\delta F$ is bounded from below. The $\varepsilon_1$ 
correction to the 
gradient term introduces dependence on $\varphi$ of the elastic constant
$K=1+\varepsilon_1{\varphi}^2$. The $\varepsilon_2$ and $\varepsilon_3$ 
terms are corrections to the potential $V_0(\varphi)$. The perturbations 
(3) are vanishing  in the disordered phase ($\varphi = 0$), while in the 
ordered phase ($\varphi\neq 0$) they may give finite contributions. 

Our paper is organised as follows. In Section 2 we 
construct the perturbative expansion. In Section 3 we apply it to the 
model (1), (2) with the particular perturbation $\delta F$ given by 
formula (3). Summary and remarks are presented in
Section 4. In the Appendix we show that the perturbatively calculated 
velocity of the planar interface coincides with 
the one calculated from an exact formula, provided that the perturbation 
is not too strong and that the perturbative series is convergent. 

\section{General structure of the perturbative expansion}

We will consider Ginzburg-Landau models in which time evolution of the order 
parameter $\varphi$ is governed by equation of dissipative type, namely 
\be
\frac{\partial\varphi}{\partial t}=\Delta\varphi-V'(\varphi,\nabla\varphi),
\ee
where $V=V_0+ \varepsilon V_1$, $t$ denotes a rescaled dimensionless time, 
$\Delta=\partial_{x^{\alpha}}\partial_{x^{\alpha}}$ 
is the three-dimensional Laplacian, 
and 
\bd 
V'=\frac{\partial V}{\partial\varphi} -
\nabla\left(\frac{\partial V}{\partial(\nabla\varphi)}\right).
\ed
We have assumed that the perturbation $V_1$ does not depend on second and 
higher derivatives of $\varphi$. 
The l.h.s. of Eq. (4) is equal to the variational derivative 
$\delta F/\delta \varphi$ multiplied by $-1$.

The interface is identified with a solution of the evolution equation which 
interpolates between minima $\varphi_{+}, \varphi_{-}$ of the potential $V$. 
Potential $V_0$ given by formula (2) has two minima, and the 
perturbation $V_1$ is by assumption weak enough, 
so that also $V$ has just two 
minima which smoothly merge with the minima of $V_0$ when $V_1$ is 
switched off. In the present paper we concentrate on planar interfaces. Due 
to translational invariance of Eq.(4), it is sufficient to consider the 
planar interfaces parallel to the plane $(x^1, x^2)$ - then $\varphi$ 
is a function of $x^3$ and 
$t$. We require that in the limits 
$x^3\rightarrow\pm\infty$,
\be
\varphi (x^3,t)\rightarrow\varphi_{\pm}, 
\ \ \ \ \ \partial_{x^3}\varphi\rightarrow 0.
\ee

Let us 
multiply Eq.(4), in which now $\nabla\varphi\rightarrow\partial_{x^3}
\varphi (x^3,t)$, by $\partial_{x^3}\varphi$ and integrate it over $x^3$ in 
the interval $(-\infty,+\infty)$. With the help of integration by parts, 
and using a formula for the full derivative of $V$ with respect to $x^3$, 
namely
\bd
\frac{dV}{dx^3}=\frac{\partial V}{\partial\varphi}\partial_{x^3}\varphi
+\frac{\partial V}{\partial(\partial_{x^3}\varphi)}\partial^{2}_{x^3}
\varphi ,
\ed
we obtain the following relation
\be
\int^{+\infty}_{-\infty} dx^3 \partial_{t}\varphi\partial_{x^3}
\varphi=V(\varphi_{-})-V(\varphi_{+}).
\ee
It implies that $\varphi$ has to be time-dependent if the minima 
$\varphi_{\pm}$ are nondegenerate. The simplest way to ensure that the l.h.s.
of formula (6) is constant in time and nonvanishing consists in assuming 
that 
$\varphi$ depends only on the combination 
$x^3-\upsilon t$, 
\be
\varphi=\varphi(x^3-\upsilon t),
\ee
where $\upsilon$ is a constant equal to the velocity of the interface. 
This form of dependence on time is also consistent with boundary 
conditions (5). For such $\varphi$, relation (6) can be written in the 
form
\be
\upsilon=\frac{V(\varphi_{+})-
V(\varphi_{-})}{\int^{+\infty}_{-\infty}dz(\partial_{z}\varphi)^2} ,
\ee
where $z=x^3-\upsilon t$. Equation (4) is reduced to
\be
\partial^{2}_{z}\varphi+\upsilon\partial_{z}\varphi-
V'(\varphi,\partial_{z}\varphi)=0 ,
\ee
where $\varphi$ is a function of $z$ only, and
\be
\varphi(z)\rightarrow\varphi_{\pm}
\ee
when $z\rightarrow\pm\infty$, respectively. The asymptotic values 
$\varphi_{\pm}$ are determined from the equations
\be
V'(\varphi_{\pm})=0.
\ee
From a mathematical viewpoint (8) and (9) should be regarded as a set of 
equations for $\upsilon$ and $\varphi(z)$, with (10) as the boundary 
conditions for $\varphi(z)$.

The perturbative Ansatz for $\varphi(z)$ has the form
\be
\varphi(z)=\varphi_0(z)+\varepsilon\varphi_1(z)+{\varepsilon}^2\varphi_2(z)
+\cdots .
\ee
Here $\varphi_0(z)$ is the initial, unperturbed interface. It obeys the 
following equation
\be
\partial^{2}_{z}\varphi_0 +\upsilon_0 \partial_{z}\varphi_0 - 
V'_0(\varphi_0)=0 ,
\ee
with the boundary conditions at $z\rightarrow\pm\infty$

\bd
\varphi_0(z)\rightarrow a_{\pm} ,
\ed
where $a_{\pm}$ are the minima of $V_0$. The velocity $\upsilon_0$ of the 
unperturbed interface is given by the formula analogous to (8), namely
\be
\upsilon_0 = \frac{V_0(a_{+})-V_0(a_{-})}{\int^{+\infty}_{-\infty} 
dz(\partial_{z}\varphi_0)^2}.
\ee
Formula (8) suggests that the velocity of the interface has the following 
perturbative expansion
\be
\upsilon=\upsilon_0+\varepsilon\upsilon_1+{\varepsilon}^2\upsilon_2+\cdots .
\ee 

Let us stress that in equations (8), (9) and in expansion (12) the 
independent variable is just $z$, and not $x^3$ and $t$. Therefore, when 
solving Eqs. (8), (9), the variable $z$ is not treated as $x^3-\upsilon t$, 
and therefore we do not expand $z$ in $\varepsilon$, contrary to what 
formula (15) might suggest. Only after $\varphi(z)$ and $\upsilon$ are 
determined we may substitute $z=x^3-\upsilon t$ in order to relate the 
solution of the set of equations (8), (9) with the interface solution of the 
original evolution equation (4).

Inserting the expansions (12), (15) in Eq.(9) and equating to zero the 
l.h.s. of (9) order by order in $\varepsilon$, we obtain the infinite chain 
of equations for the corrections $\varphi_{k}, k\geq 1$. These equations 
have the form 
\be
\hat{L}\varphi_{k}=f_{k} ,
\ee
where
\be
\hat{L}={\partial}^{2}_{z}+\upsilon_0\ \partial_{z}-V_{0}''(\varphi_0) ,
\ee
and $f_{k}$ in general depend on $\upsilon_1,\ldots, \upsilon_{k}$ and 
$\varphi_1,\ldots, \varphi_{k-1}$. Simple calculations give
\be
f_1=-\upsilon_1\ \partial_{z}\varphi_0 + V_{1}'(\varphi_0, 
\partial_z \varphi_0) ,
\ee
\noindent 
and for $k\geq 2$

\be
f_{k}=-\upsilon_{k}\partial_{z}\varphi_0-
\sum^{k-1}_{j=1}\upsilon_{j}\ \partial_{z}\varphi_{k-j} + h_{k}(z) ,
\ee
where $h_{k}(z)$ denotes the sum of all terms of the $k$-th order in 
$\varepsilon$ obtained by inserting formula (12) in $V_{0}'(\varphi)+
\varepsilon V_{1}'(\varphi,\partial_z\varphi)$ with exception of the term 
$V_{0}''(\varphi_0)\varphi_{k}$ which has been included into the l.h.s. of 
Eq.(16).

It is a well-known fact that inhomogeneous linear equations not always have 
solutions - certain integrability conditions have to be satisfied. In the 
case of Eq.(16) such conditions appear because for the operator $\hat{L}$ 
there exists so called left zero-mode, that is a normalizable 
function $\psi_{l}(z)$ such that the following identity holds
\be
\int^{+\infty}_{-\infty} 
dz \psi_{l}(z)\hat{L}\varphi_{k}=0.
\ee
If $\varphi_{k}$ obeys Eq.(16) then
\be
\int^{+\infty}_{-\infty} dz \psi_{l}(z)f_{k}=0.
\ee
In our perturbative scheme the integrability conditions (21) serve as 
equations which determine the perturbative contributions $\upsilon_{k}$ to 
the velocity of the interface. Using formulas (18), (19) and conditions (21) 
we find that 
\be
\upsilon_1=N^{-1}\int^{+\infty}_{-\infty} dz V_{1}'(\varphi_0, 
\partial_z \varphi_0)\psi_{l}(z) ,
\ee
and for $k\geq 2$
\be
\upsilon_{k}=N^{-1}\int^{+\infty}_{-\infty} dz
\left(h_{k}(z)-\sum^{k-1}_{j=1}\upsilon_{j}\ 
\partial_{z}\varphi_{k-j}\right)\psi_{l}(z) ,
\ee
where
\bd
N=\int^{+\infty}_{-\infty} dz\psi_{r}(z)\psi_{l}(z).
\ed
The functions $\psi_{l}, \psi_{r}$ will be given shortly. In the Appendix 
we check that these recursive formulas and the expansion (15) give the 
velocity $\upsilon$ which coincides with the one obtained by expanding in 
$\varepsilon$ the r.h.s. of formula (8), provided that $\varepsilon$ is not 
too large.

The left zero-mode $\psi_{l}$ can be found in the 
following way. Differentiation of the both sides of Eq.(13) with respect to 
z gives the identity
\be
\hat{L}\psi_{r}=0 ,
\ee
where
\be
\psi_{r}=\partial_{z}\varphi_{0}(z).
\ee
The function $\psi_{r}$ is called the right zero-mode. $\psi_{l}$ has the 
form
\be
\psi_{l}(z)=\exp(\upsilon_0 z)\psi_{r}(z).
\ee
Because derivation of identity (20) involves integration by parts, we have to 
discuss behaviour of $\psi_l$ and $\psi_r$ for $z\rightarrow\pm\infty$. 
Notice that Eq.(24) implies that for $z\rightarrow\pm\infty$
\bd
\psi_{r}(z)\cong\exp\left(-\frac{\upsilon_0}{2} z\right)
\exp\left(\mp\sqrt{\frac{{\upsilon_0}^2}{4}+V_{0}''(a_{\pm})} \ 
\ z\right) ,
\ed
respectively. Therefore $\psi_{l}(z)$ exponentially vanishes for 
$z\rightarrow\pm\infty$, provided that 
\bd
V_{0}''(a_{\pm})>0.
\ed
For $V_0$ given by formula (2) this condition is satisfied if 
$0<A<\frac{9B^2}{32}$, as it has been assumed. When checking identity (20),
boundary terms like, e.g., 
$\left.(\partial_{z}\psi_{l})\varphi_{k}\right|^{+\infty}_{-\infty}$, 
vanish because
\be
\varphi_{k}\rightarrow \mbox{const} \ \ \ \  \mbox{for} 
 \ \ \ \ z\rightarrow\pm\infty  ,
\ee
in accordance with the boundary condition (10).

Equations (16) can be solved in a standard way \cite{8}. Adopting 
formulas given in \cite{8} to the case at hand we find that 
\be
\varphi_{k}=\psi_{2}(z)\int^{z}_{-\infty}d\zeta 
e^{\upsilon_0\zeta}\psi_{r}(\zeta)f_{k}(\zeta) - 
\psi_{r}(z)\int^{z}_{0}d\zeta e^{\upsilon_0\zeta}\psi_{2}(\zeta)f_{k}(\zeta) ,
\ee
where
\be
\psi_{2}(z)=\psi_{r}(z)\int^{z}_{0} d\zeta 
e^{-\upsilon_0\zeta}{\psi}^{-2}_{r}(\zeta).
\ee
The functions $\psi_{r}, \psi_{2}$ form a pair of linearly independent 
solutions of the homogeneous equation $\hat{L}\varphi=0$. Notice that the 
integral $\int^{z}_{-\infty}$ present in the first term on the r.h.s. of 
formula (28) vanishes for $z\rightarrow +\infty$ due to the integrability 
conditions (21). It turns out that in the limits $z\rightarrow\pm\infty$ 
both terms on the r.h.s. of formula (28) approach constants, 
and $\varphi_{k}$ obey the conditions (27).

General solution of Eqs.(16) is obtained by adding to the r.h.s. of 
formula (28) the general solution of the corresponding 
homogeneous equation, that is the function $c_{k}\psi_{r}(z)+d_{k}
\psi_{2}(z)$. However, because $\psi_{2}(z)$ exponentially grows for large 
$|z|$ we have to put $d_{k}=0$. In order to determine the constants $c_{k}$ 
we have to impose a condition on the interface solution $\varphi(z)$ in 
addition to the boundary conditions (10). Actually, Eq.(9) and the boundary 
conditions (10) do not determine $\varphi(z)$ uniquely - due to 
invariance with respect to translations in $z$ we can take $\varphi(z-z_0)$ 
with arbitrary $z_0$. We will require that
\be
\varphi(0)=\varphi_0(0),
\ee
where $\varphi_0(z)$ is a concrete, explicitly given 
function describing the 
unperturbed interface with fixed location on the $z$ axis. The condition 
(30) breaks the translational invariance. It implies that
\be
\varphi_{k}(0)=0.
\ee
The solutions $\varphi_{k}(z)$ given by formula (28) obey this condition, 
therefore also the constants $c_k$ are vanishing. Thus, the perturbative 
scheme supplemented with the condition (30) yields unique interface solution 
$\varphi(z)$. In this solution we can of course shift the variable $z$, 
that is to substitute $z\rightarrow z-z_0$ simultaneously in all 
contributions $\varphi_0, \varphi_k$, in 
order to obtain the other solutions implied by the translational invariance.

\section{Effects of the perturbations of the form \\
$\;\;\;\;\;\;V_1=\frac{1}{2}\varepsilon_1{\varphi}^2
\nabla\varphi\nabla\varphi+
\frac{1}{5}\varepsilon_2{\varphi}^5+\frac{1}{6}\varepsilon_3{\varphi}^6$.}

Let us start from a description of the unperturbed interface $\varphi_0(z)$. 
The evolution equation has the form (4) with $V$ replaced by $V_0$ given by
formula (2). The interface solution is well-known, see, e.g., \cite{4}. 
Let us quote the relevant formulas from \cite{4} and \cite{6}. We use the 
following abbreviations
\bd
s=\frac{z}{4 l_0}, \ \ \ \ \ \ \ \ 
\  a_{+}=-\frac{1}{2\sqrt{2} l_0} ,
\ed
where 
\be
{l_0}^{-1}=\frac{1}{2\sqrt{2}}\left(3B+\sqrt{9B^2-32A}\right).
\ee
Then
\be
\varphi_0(z)=\frac{a_{+}}{2}\left(1+\tanh{s}\right) ,
\ee
and
\beq
\psi_{r}(z) &=& \frac{a_{+}}{8l_0}\frac{1}{\cosh^2 s} \nonumber  , \\
\psi_2(z) &=& \frac{2 {l_0}^2}{a_{+}} \frac{1}{\cosh^2 s} \int_0^s d\sigma
\exp[-2(2\alpha+1)\sigma] \left[1+ \exp(2\sigma)\right]^4. \nonumber
\eeq
The integral over $\sigma$ can be easily calculated, 
but the resulting formula for $\psi_2$ is quite long. 
In the absence of perturbations, $z=x^3-\upsilon_0 t$, 
where
\be
\upsilon_0=\frac{3}{4\sqrt{2}}\left(B-\sqrt{9B^2-32A}\right).
\ee
\noindent 
The interface $\varphi_0$ exists and it is stable when $0<A<\frac{9B^2}{32}$.
Usually, the coefficient $B$ is regarded as independent of temperature, while 
$A=a(T-T_{\ast})$, where $a>0$. The inequality given above fixes the 
temperature range $(T_{\ast},T_{c})$ in which the interface exists. Notice 
that at the temperature $T_0$ such that $A=\frac{B^2}{4}$ the velocity 
$\upsilon_0$ vanishes. It is clear that $T_{\ast}<T_0<T_{c}$. The left 
zero-mode has the form
\be
\psi_{l}(z)=\frac{a_{+}}{8l_0}\exp{[2(2\alpha-1)s]}\frac{1}{\cosh^2 s} ,
\ee
where
\bd
\alpha=\upsilon_0 l_0+\frac{1}{2}.
\ed
The value of the parameter $\alpha$ monotonically increases from 0 to 1 when 
$A$ varies from 0 to $\frac{9B^2}{32}$.

The first order correction $\upsilon_1$ to the velocity is calculated from 
formula (22). In the present case
\be
V_1'(\varphi_0)=-\varepsilon_1\varphi_0{\psi_{r}}^2-
\varepsilon_1{\varphi_0}^2\partial_{z}\psi_{r}+
\varepsilon_2{\varphi_0}^4+\varepsilon_3{\varphi_0}^5.
\ee
The integrations over $z$ ( or equivalently over $s$ ) can be related to 
the Euler $B$ function \cite{9},
\bd
B(x,y)=\int^{\infty}_{0} dt \frac{t^{x-1}}{(1+t)^{x+y}} ,
\ed
by the substitution $y=\exp{s}$. The resulting expressions can be 
simplified with the help of the well-known formulas
\bd
B(x,y)=\frac{\Gamma (x)\Gamma (y)}{\Gamma (x+y)}, \ \ \ \ \ \ \ \ \ \
\Gamma (x+1)=x\Gamma (x).
\ed
After straightforward calculations we obtain
\be
\upsilon_1=\frac{(\alpha+1)(2\alpha+1)}{160 {l_0}^3}
\left[(\alpha-\frac{1}{2})\varepsilon_1-
\frac{l_0}{\sqrt{2}}\frac{2\alpha+3}{1-\alpha}\varepsilon_2+
\frac{(2\alpha+3)(\alpha+2)}{12(1-\alpha)}\varepsilon_3\right].
\ee

Formula (37) shows that the $\varepsilon_2, \varepsilon_3$ corrections 
become more and more pronounced as $\alpha$ increases towards $1$, 
that is as the temperature increases towards $T_{c}$. The singularity 
at $\alpha=1$ is due to the fact that for $\alpha=1$ the left zero-mode 
$\psi_{l}$ and $V_1'(\varphi_0)$ do not vanish when $s\rightarrow +\infty$, 
and therefore the integral in formula (22) is divergent.

The $\varepsilon_1$ term in formula (37) vanishes for $\alpha=\frac{1}{2}$. 
The reason is that the corresponding term in the perturbation potential 
$V_1$ does not contribute to the values of the full potential $V$ at the 
two minima $\varphi_{\pm}$. Therefore, it can influence the velocity only
by changing the denominator in formula (8). For $\alpha=\frac{1}{2}$ the 
numerator in (8) vanishes if $\varepsilon_2=\varepsilon_3=0$ and the 
$\varepsilon_1$ term can not contribute to the velocity.

The facts that $\varepsilon_2$ term in (37) is negative, and that the 
$\varepsilon_3$ term is positive, can be explained by the observation that 
the $\varepsilon_2$ term in the potential diminishes the potential energy 
difference across the interface, while the $\varepsilon_3$ term in the 
potential works in the opposite direction.

\section{Summary and discussion}

1. The purpose of this work is to provide a practical tool for analytic, 
approximate computations of characteristics of the perturbed planar
interfaces within the framework of Ginzburg-Landau models. We have shown
how one can systematically calculate corrections to the profile and the
velocity of the interface. In particular, the scheme can be applied to
perturbations of domain walls, which can be regarded as special, static
interfaces appearing when the two minima of $V_0$ are degenerate. 
In the example discussed in Section 3 this is the case when $\alpha=1/2$.
The scheme is relatively simple. The final formulas (22), (23), 
(28), (29) contain
one dimensional integrals which always can be tackled numerically, and 
in many cases calculated analytically.  

\noindent
2. We have not delved into the problem of convergence of the perturbative
series. Such a mathematical investigation does not belong to the scope of
our work. It is clear that if the perturbations shift the two minima of
$V_0$ only slightly, the corresponding changes of 
the profile and of the velocity of the
interface are also small, hence they can be calculated as small corrections. 
This does not necessarily mean that the series is convergent -- it can
belong to a wider class of asymptotic series, but it is acceptable from
physical viewpoint. 

\noindent
3. Using the results of the present paper in combination with the 
curvature expansion mentioned in the Introduction, one can  calculate
evolution of curved interfaces in the perturbed Ginzburg-Landau models.
Comparisons of such theoretical predictions with experimental observations of 
evolution of the interfaces could help to determine the best formula for
the free energy of the system.  Let us remind that the formulas for the 
free energy in the Ginzburg-Landau models rarely can be derived from 
underlying microscopic theories. In most cases they are postulated on basis
of qualitative phenomenological considerations. General discussion of
various formulas for the free energy in the context of theory of liquid 
crystals can be found in \cite{10}.

\section{Appendix. Resummation of the perturbative contributions to the 
velocity }

The perturbative corrections $\upsilon_{k}$ to the velocity of the 
interface are given by formulas (22), (23). They have been obtained in 
somewhat indirect way, namely from the integrability conditions. On the 
other hand, the velocity is given by the exact formula (8). We would like to 
check that the sum (15) of the perturbative contributions $\upsilon_{k}$, 
below denoted by $\tilde{\upsilon}$,
\be
\tilde{\upsilon}=\!\!\!\!\!\!^{df} \upsilon_0+\varepsilon\upsilon_1+
\sum^{\infty}_{k=2}{\varepsilon}^k\upsilon_{k} ,
\ee
does coincide with the velocity $\upsilon$ given by formula (8). 
Actually, because formula (8) follows directly from Eq.(9), it is sufficient 
to check that $\tilde{\upsilon}$ coincides with the $\upsilon$ present in 
that equation.

In the first step, we substitute for $\upsilon_0, 
\upsilon_1$ and $\upsilon_{k}$ ( $k\geq 2$ ) in (38) formulas (14), (22) 
and (23), respectively. Next, we notice that the definition of $h_{k}$ given 
below formula (19) is equivalent to the following formula
\bd
\sum^{\infty}_{k=2}{\varepsilon}^k h_{k}=V_0'(\varphi)+
\varepsilon V_1'(\varphi, \partial_{z}\varphi)-V_0'(\varphi_0)-
\varepsilon V_1'(\varphi_0, \partial_{z}\varphi_0)-
V_0''(\varphi_0)(\varphi-\varphi_0) ,
\ed
where
\bd
V_0'(\varphi)+\varepsilon V_1'(\varphi, \partial_{z}\varphi)
={\partial_{z}}^2\varphi+\upsilon\partial_{z}\varphi ,
\ed
\bd
V_0'(\varphi_0)=\partial_{z}\psi_{r}+
\upsilon_0\psi_{r},
\ed
and
\bd
V_0''(\varphi_0)\varphi_0=\partial_{z} \psi_r+\upsilon_0\psi_r-
\hat{L}\varphi_0 ,
\ed
according to Eqs. (9), (13) and definition (17). We also use the formula
\bd
\sum^{\infty}_{k=2}{\varepsilon}^{k}\sum^{k-1}_{j=1}\upsilon_{j}
\partial_{z}\varphi_{k-j}=\left(\sum^{\infty}_{i=1}{\varepsilon}^{i}
\upsilon_{i}\right)\left(\sum^{\infty}_{j=1}{\varepsilon}^{j}\partial_z
\varphi_{j}\right)=(\tilde{\upsilon}-\upsilon_0)\partial_z(\varphi-\varphi_0).
\ed
After a simple calculation we obtain from formula (38) that
\be
(\upsilon-\tilde{\upsilon})\int^{+\infty}_{-\infty}dz \psi_{l}\partial_{z}
\varphi =0.
\ee
Thus, indeed
\bd
\upsilon=\tilde{\upsilon} ,
\ed
provided that
\bd
\int^{+\infty}_{-\infty}dz \psi_{l}\partial_{z}\varphi \neq 0.
\ed
Because 
\bd
\partial_{z}\varphi=\psi_{r}+\varepsilon\partial_{z}\varphi_1+\cdots ,
\ed
that last condition is satisfied if $\varepsilon$ is small enough, that is 
if the perturbation is not too strong.

\end{document}